\documentclass[preprint,aps,showkeys,showpacs]{revtex4}
\usepackage[english]{babel}
\usepackage{graphicx}
\usepackage{graphics}
\usepackage{amsmath}
\usepackage{dcolumn}
\usepackage{amssymb}
\usepackage{bm}

\begin{document}

\title{Lorentz violation and higher-derivative gravity}

\author{C. A. Hernaski}
\email{chernask@indiana.edu}
\affiliation{Indiana University Center for Spacetime Symmetries,
Bloomington, Indiana 47405, USA}
\author{H. Belich}
\email{belichjr@gmail.com}
\affiliation{Departamento de F\'{\i}sica e Qu\'{\i}mica, Universidade Federal do Esp\'{\i}rito Santo, Avenida Fernando Ferrari, 514, Goiabeiras, 29060-900, Vit\'{o}ria,
ES, Brazil.}

\begin{abstract}
In this work, we analyze a gravity model with higher derivatives including a
CPT-even Lorentz-violating term. In principle, the model could be a low-energy limit of a Lorentz-invariant theory presenting the violation of
Lorentz symmetry as a consequence of a spontaneous symmetry-breaking
mechanism if a decoupling between the metric and the Nambu-Goldstone modes
is assumed. We have set up a convenient operator basis for the expansion
of wave operators for symmetric second-rank tensors in the presence of a
background vector. By using this set of operators, the particle content is
obtained, and its consistency, regarding the conditions for stability and unitarity, is discussed. We conclude that this extra Lorentz
noninvariant contribution is unable to address the problems of stability
and unitarity of higher-derivative gravity models.
\end{abstract}

\keywords{Lorentz Breaking, Gravity Models, Higher Derivatives}
\pacs {04.60.-m, 04.50.Kd, 11.30.Er}
\maketitle

\section{Introduction}

Problems involving the consistent quantization of the gravitational
interaction have been widely discussed and are well known. The standard
quantization of the Einstein-Hilbert (EH) Lagrangian using the formalism of
perturbative quantum field theory (QFT) turns out to be inadequate, since it
results in a non-renormalizable model \cite{weinberg}. Attempts to formulate a consistent
quantum gravity model in four spacetime dimensions led to the proposal of
additional terms involving higher derivatives. These terms modify the
propagating structure of the excitations (including the graviton) in order
to improve the ultraviolet (UV) behavior of the scattering amplitudes \cite{neville1,neville2,stelle}.
However, invariably, they also introduce negative-norm excitations
called ghost particles that damage the unitarity of the S matrix. As
discussed in Ref. \cite{new}, this is a general feature of gravitational models
built up with diffeomorphism-covariant terms and points to an
incompatibility between renormalizability and unitarity of the S matrix.

Many attempts to try to circumvent this incompatibility were discussed in the
literature \cite{neville1,neville2,stelle,new,lite}. An interesting possibility that evades some previous assumptions
is to consider the Lorentz and diffeomorphism-symmetry-breaking effects on
these gravity models, which, in turn, should describe the phenomenological
aspects of general relativity (GR) in a low-energy limit.

The interest in models that present the breaking of the Lorentz symmetry has
increased after Kosteleck\'{y} and Samuel showed that string
theory, one of the leading candidates to handle the issue of consistent
quantization of gravity, may present some phases where Lorentz symmetry is
violated through a spontaneous symmetry-breaking mechanism triggered by the
appearance of nonvanishing expectation vacuum values of nontrivial Lorentz
tensors \cite{extra3}. In this case, the models with Lorentz symmetry
breaking are considered effective, and the analysis of their phenomenological
aspects at low energies may provide information and impose restrictions on
the fundamental theory from which they stem.

A general framework for testing the low-energy manifestations of CPT and
Lorentz symmetry breaking is the Standard-Model Extension (SME) \cite{colladay-kost}.
In this framework, the effective Lagrangian corresponds to
the usual Lagrangian of the Standard Model (SM), to which is added SM
operators of any dimensionality contracted with Lorentz-violating (LV) tensorial
background coefficients. The effective Lagrangian is written in a Lorentz-invariant form under coordinate transformations to guarantee the
observer independence of physics. However, the physically relevant
transformations are those that affect only the dynamical fields of the
theory. These changes are called particle transformations, whereas the
coordinate transformations (including the background tensor) are called
observer transformations. In Refs. \cite{coll-kost,bras}, these
concepts are more deeply analyzed.

To contemplate the possible consequences of CPT and
LV for experiments where the gravitational interaction plays
an important role, gravity was also included in the general framework of the SME 
\cite{alan-gravity}. One of the most striking consequences of this fusion is
the requirement of a dynamic mechanism for the diffeomorphism/Lorentz-invariance violation \textemdash that is, the violation of these spacetime symmetries
should be spontaneous to be consistent with the geometrical aspects
of the gravitational interaction \cite{finsler}. In practice, this means that the tensor
coefficients appearing in the diffeomorphism/Lorentz-violating operators
must be replaced by dynamical fields, and these, in turn, under a specific
dynamic mechanism, acquire nonvanishing vacuum expectation values. As
compared with the scenario of explicit breaking, one finds extra
Nambu-Goldstone and massive modes that are responsible for restoring the
particle diffeomorphism/Lorentz invariance of the model \cite{fung,bluhm}.

Concerning the experimental searches for the CPT/Lorentz-violation signals, the generality of the SME
has provided the basis for many investigations. In the flat spacetime limit, the 
empirical studies include muons \cite{muon}, mesons \cite{meson,meson2}, 
baryons \cite{barion}, photons \cite{photon}, electrons \cite{electron}, 
neutrinos \cite{neutrino}, and the Higgs \cite{higgs} sector. The gravity sector was also 
explored in Refs. \cite{gravity,torsion,tasson,gravity2}. In Ref. \cite{data}, one can find the current limits on 
coefficients for Lorentz violation.

The aim of this work is to investigate if, by considering a
diffeomorphism/Lorentz-violating CPT-even term, one can get a model with the
spectrum free from instabilities and, at the same time, accompanied by
higher derivatives, since an improved UV behavior of the scattering
amplitudes is expected in the presence of such terms. For the discussion of these
issues, we have analyzed the properties of the particle propagators that
show up in this model. For this task, we have concentrated on the quadratic
Lagrangian resulting from the expansion of the dynamical fields around a
vacuum solution to the equations of motion. For simplicity, we have decided 
to focus only on the propagating modes that come from the dynamical
metric, leaving aside the Nambu-Goldstone and massive modes that should
appear as the result of the spontaneous symmetry-breaking mechanism. This
choice relies on the fact that one can write interacting
Lagrangians where these extra modes decouple from the metric modes, and one
can study the properties of the latter independently of the former. However,
one should be aware that the full consistency of the interacting system
should include these extra modes. We hope that this inconsistency does not
appear perturbatively at the quadratic level, but this possibility should
be explicitly investigated \cite{manoel}. Assuming the validity of such a decoupling, we
work out the propagators of the model using a suitable operator basis for
the wave operator that highlights the physical degrees of freedom (d.o.f.)
propagating in the model, as well as keeping track of the role of the Lorentz
breaking in providing extra couplings between these d.o.f. In the following,
we perform a general analysis of the stability and unitarity, pointing out
the extra subtleties brought by the LV. Our conclusion is that
the extra room obtained by removing the Lorentz invariance is not
enough to render the higher-derivative model consistent. Unless we consider
the model from a effective point of view and assume some parameters of the
Lagrangian to be of the same order as some assumed cutoff, the model has
insurmountable consistency difficulties even in the infrared regime.

This paper is organized as follows: In Sec. \ref{section 2}, we set up the
notation and conventions to be used, and we start with a general Lagrangian
containing the above mentioned LV term in addition to the EH and curvature-square terms. The collection and analysis of propagators is carried out in
Sec. \ref{subsection 2.1}. The conditions for stability of the spectrum are attained
in Sec. \ref{subsection 2.2}. In Sec. \ref{section 3}, we present our discussions and conclusions.

\section{Gravitation with higher derivatives and violation of Lorentz
symmetry \label{section 2}}

In Ref. \cite{prdhernaski2011} is carried out an analysis of the spectral consistency of a
gravitational model in $3+1$ dimensions that contains, in addition to the
EH term, terms with higher derivatives $(R^{2}$ and $R_{ab}R^{ab})$ plus two
CPT-odd terms which manifest the breaking of the CPT and
diffeomorphism/Lorentz symmetry through a constant background vector: a
Chern-Simons-like and a Ricci-Cotton-like term. As in the case of
gravitation without diffeomorphism/Lorentz breaking, we concluded that the higher-derivative
terms should be avoided to get a stable model. The only consistent combination is the EH Lagrangian
added to the Chern-Simons-type term with a timelike background vector. In this
work, we address the analysis of the compatibility of unitarity with higher-derivative terms by considering a CPT-even diffeomorphism/Lorentz-violating
operator. We consider the following Lagrangian:
\begin{equation}
\mathcal{L}=\sqrt{-g}\left( \alpha R+\beta R_{ab}R^{ab}+\gamma R^{2}+\frac{1%
}{4}\kappa ^{abcd}R_{abfg}R_{\ \ cd}^{fg}\right) .  \label{principal}
\end{equation}
The parameters $\alpha $, $\beta $, and $\gamma $ are arbitrary, and the
background tensor $\kappa ^{abcd}$ is responsible for the breakdown of the
diffeomorphism invariance that, whenever we consider the linearized model,
will lead to the breaking of Lorentz symmetry. To simplify the
discussion, we will make the additional assumption that this tensor is
constructed with a background vector $b^{a}$ \cite{extra3}. Then, we have
\begin{equation}
\kappa ^{abcd}=\frac{1}{2}\left( \kappa ^{ac}g^{bd}+\kappa
^{bd}g^{ac}-\kappa ^{bc}g^{ad}-\kappa ^{ad}g^{bc}\right) ,  \label{fundo}
\end{equation}%
with
\begin{equation}
\kappa ^{ab} =\kappa \left( b^{a}b^{b}-\frac{1}{4}g^{ab}b\cdot b\right)
\end{equation}
and 
\begin{equation}
\kappa =\frac{4}{3}\kappa ^{ab}b_{a}b_{b}.  \label{dois}
\end{equation}
The conventions adopted for the Riemann tensor, Ricci and scalar curvature
are those followed by Ref. \cite{misner}. That is,
\begin{eqnarray}
R_{\ bcd}^{a} &=&\partial _{c}\Gamma _{bd}^{a}-\partial _{d}\Gamma
_{bc}^{a}+\Gamma _{ce}^{a}\Gamma _{bd}^{e}-\Gamma _{de}^{a}\Gamma _{bc}^{e},
\\
R_{bc} &=&R_{\ bac}^{a},\ \ \ \ R =g^{ab}R_{ab},
\end{eqnarray}%
and the Christoffel symbol $\Gamma _{bc}^{a}$ is given by
\begin{equation}
\Gamma _{bc}^{a}=\frac{1}{2}g^{ad}\left( \partial _{b}g_{dc}+\partial
_{c}g_{db}-\partial _{d}g_{bc}\right) .
\end{equation}
The background tensor given in Eq. (\ref{fundo}) has the same symmetries as the
Riemann tensor, as can easily be seen by the structure of the breaking
term in Eq. (\ref{principal}). We can assume that the origin of the background
vector, $b^{a}$, is due to a spontaneous breaking of Lorentz symmetry in a
more fundamental theory, as for example in string theory \cite{extra3}. The extra modes coming from the spontaneous
symmetry-breaking mechanism are not discussed, and we expect that for some
classes of models this can be done independently without disturbing the
present analysis, as discussed in the Introduction. Furthermore, the nature
of the background vector, if it is space-, time-, or lightlike, is not
assumed {\it a priori}. We discuss the consistency of the model, case by case, in
the following sections. The values of the constants appearing in Eq. (\ref{principal})
must be prescribed in such a way to obtain a model with a particle
content free of ghosts and tachyons.

The Minkowski metric corresponds to a possible solution of the
Euler-Lagrange equations obtained from Eq. (\ref{principal}). Therefore, we can
consider a perturbation of the metric field around the Minkowski metric,
\begin{equation}
g_{ab}=\eta _{ab}+h_{ab},
\end{equation}%
where $\eta _{ab}=diag\left( 1,-1,-1,-1\right)$. In terms of $h_{ab}$, the
Lagrangian [Eq. (\ref{principal})] establishes the dynamics for this field in an
anisotropic spacetime scenario. Furthermore, we will assume that the background vector is constant in the asymptotic Minkowski coordinates
\begin{equation}
\partial_{a}b^c=0. \label{constant back}
\end{equation}
We should highlight that this condition is not equivalent to
the covariant constancy of the background vector, $b_{\mu}$, under general diffeomorphisms. The (covariant) constancy of $b_{\mu}$
does not hold for arbitrary manifolds; it rather imposes a constraint
on the curved space \cite{alan-gravity}. For that reason, we are not assuming $b_{\mu}$
to be covariantly constant. We relax this more stringent condition and, in simply considering metric fluctuations (weak-field approximation), it is legitimate to consider
constancy in the asymptotic Minkowskian sense, as stated above in Eq. (\ref{constant back}).

The particle content described by this model can be examined considering the propagators, whose structures depend
only on the quadratic Lagrangian in $h_{ab}$. Up to total derivatives, it is given by
\begin{eqnarray}
\left( \mathcal{L}\right) _{2} &=&\frac{\alpha }{2}\left( \frac{1}{2}h^{\mu \nu }\square h_{\mu \nu }-\frac{1}{2}
h\square h+h\partial _{\mu }\partial _{\nu }h^{\mu \nu }-h^{\mu \nu}\partial _{\mu }\partial _{\lambda }h_{\ \nu }^{\lambda }\right)  \notag \\
&+&\frac{\beta }{2}\left( h\square ^{2}h-2h\square\partial _{\mu }\partial _{\nu }h^{\mu \nu }+h^{\mu \nu }\partial _{\mu
}\partial _{\nu }\partial _{\kappa }\partial _{\lambda }h^{\kappa \lambda}\right)  \notag \\
&+&\frac{\gamma }{8}\left(h^{\mu \nu }\square ^{2}h_{\mu\nu }-2h\square \partial _{\mu }\partial _{\nu }h^{\mu \nu }+2h^{\mu \nu}
\partial _{\mu }\partial _{\nu }\partial _{\kappa }\partial _{\lambda}h^{\kappa \lambda }+h\square ^{2}h\right)    \notag \\
&+&\frac{\kappa b^{a}b^{c}}{4}\Big(2h_{ag}\partial _{f}\partial ^{d}\partial _{c}\partial ^{g}h_{\ d}^{f}-h_{\ g}^{d}\partial _{f}\partial _{a}\partial _{c}\partial ^{g}h_{\ d}^{f}  \notag \\
&-&2h_{af}\partial ^{d}\partial _{c}\square h_{\ d}^{f}+h_{\ f}^{d}\partial _{a}\partial _{c}\square h_{\ d}^{f}-h_{ag}\partial _{f}\partial ^{g}\square h_{\ c}^{f}+h_{af}\square ^{2}h_{\ c}^{f}\Big)  \notag \\
&-&\frac{\kappa b\cdot b}{8}\left(h_{ag}\partial _{f}\partial ^{d}\partial ^{a}\partial ^{g}h_{\ d}^{f}+h_{\ f}^{d}\square ^{2}h_{\ d}^{f}-2h_{af}\partial ^{d}\partial ^{a}\square h_{\ d}^{f}\right).  \label{perturba}
\end{eqnarray}
It is important to notice that, although the background vector breaks the
diffeomorphism invariance of the Lagrangian [Eq. (\ref{principal})], the quadratic
Lagrangian has the gauge invariance
\begin{equation}
h_{ab}^{\prime }=h_{ab}+\partial _{a}\xi _{b}+\partial _{a}\xi _{b}.
\label{gauge}
\end{equation}
This is due to the fact that the model in Eq. (\ref{principal}) is constructed
using only scalar and tensor curvatures. Under general diffeomorphism
transformations, these quantities transform covariantly (or invariantly in
the case of scalars). When considering the linear contribution in the field $h_{ab}$,
these terms are invariant under the gauge transformation [Eq. (\ref{gauge})] arising 
from the linearized diffeomorphisms
\begin{equation}
x^{\prime \mu }=x^{\mu }+\xi ^{\mu }\left( x\right) .  \label{lin trans}
\end{equation}
As in GR, the gauge invariance in Eq. (\ref{gauge}) is responsible for inhibiting
the propagation of a spin-$1$ and a spin-$0$ mode \cite{weinberg2}. Moreover, dynamically, the
graviton keeps its two helicity d.o.f, as we shall see in the
next section.

\subsection{Obtaining the propagators \label{subsection 2.1}}

To characterize the particle content that propagates in this
spacetime scenario, we analyze the propagator structure of the
excitations. Then, we first write Eq. (\ref{perturba}) as follows:

\begin{equation}
\left( \mathcal{L}\right) _{2}=\frac{1}{2}h^{ab}\mathcal{O}_{ab;cd}h^{cd},
\label{onda}
\end{equation}%
where $\mathcal{O}$ is the wave operator. The task of obtaining the
propagators is equivalent to finding the inverse of this operator. For this, we
will use an algebraic method by finding an operator basis in terms of which
we can expand it. The two main criteria for the construction of this basis
will be algebraic simplicity and easy physical interpretation. As we will
see later, the first criterion is fulfilled if we build an operator basis
consisting of projectors and mappers. The vector space of the field in
question is then divided into subspaces defined by projectors, and those with the
same dimension can be mapped in each other by means of the mappers. Our
interest is in the particle content described by the model. Thus,
in order to satisfy the second criterion, we define the subspaces in such a
way as to associate a well-defined spin to the d.o.f. of the fields
that reside in these subspaces.

In models with Lorentz invariance, the kinetic terms of the Lagrangian are
built using only metrics and derivatives. Thus, the kinetic term for the
field $h^{ab}$, if it is Lorentz invariant, should be such that the wave
operator could be expanded in terms of Barnes-Rivers operators~\cite{barnes}. These, in turn, are given by
\begin{eqnarray}
P(2)_{ab;cd} &=&\frac{1}{2}(\theta _{ac}\theta _{bd}+\theta _{ad}\theta _{bc})-\frac{1}{3}\theta
_{ab}\theta _{cd}),  \label{projspin2} \\
P(1)_{ab;cd} &=&\frac{1}{2}(\theta _{ac}\omega _{bd}+\theta _{ad}\omega _{bc}+\theta _{bc}\omega
_{ad}+\theta _{bd}\omega _{ac}),  \label{projspin1} \\
P_{11}(0)_{ab;cd} &=&\frac{1}{3}\theta _{ab}\theta _{cd},  \label{projspin01} \\
P_{22}(0)_{ab;cd} &=&\omega _{ab}\omega _{cd},  \label{projspin02} \\
P_{12}(0)_{ab;cd} &=&\frac{1}{\sqrt{3}}\theta _{ab}\omega _{cd},  \label{mapspin012} \\
P_{21}(0)_{ab;cd} &=&\frac{1}{\sqrt{3}}\omega _{ab}\theta _{cd}.  \label{mapspin021}
\end{eqnarray}%
However, the presence of background tensors allows extra couplings to the fields
in such a way that the Barnes-Rivers operators are, in general,
insufficient to expand the wave operators containing these structures.

We can arrange the terms of the quadratic Lagrangian [Eq. (\ref{perturba})] into
two distinct classes. One is composed of terms in which at least one of the $h$ fields is contracted with the background vector, and the other
presents terms in which the background vectors are contracted with each
other or with the derivatives. Thus, we rewrite Eq. (\ref{onda}) as follows:
\begin{equation}
\left( \mathcal{L}\right) _{2}=\frac{1}{2}h^{ab}\mathcal{O}%
_{ab;cd}^{1}h^{cd}+\frac{1}{2}h^{ab}\mathcal{O}_{ab;cd}^{2}h^{cd},
\label{onda quebra}
\end{equation}
with
\begin{eqnarray}
\mathcal{O}_{ab;cd}^{1} &=&\frac{1}{2}\kappa p^{2}[-\left( v\cdot p\right) (v_{d}\eta
_{bc}p_{a}-v_{a}\omega _{cd}p_{b}+v_{a}\eta _{bc}p_{d}  \notag \\
&&-v_{d}\omega _{ab}p_{c})+p^{2}\left( v_{a}v_{d}\eta _{bc}-v_{a}v_{d}\omega _{bc}\right) ]+\cdots  \label{op quebra}
\end{eqnarray}
and
\begin{eqnarray}
\mathcal{O}^{2} &=&\alpha p^{2}\left(\frac{1}{2}\eta _{ac}\eta _{bd}-\frac{1}{2}\eta _{ab}\eta _{cd}-\eta
_{ac}\omega _{bd}+\eta _{ab}\omega _{cd}\right)  \notag \\
&+&\beta p^{4}\left(\frac{1}{2}\eta _{ac}\eta _{bd}+\frac{11}{6}\eta _{ab}\eta _{cd}+\frac{7}{3}\omega
_{ab}\omega _{cd}-\eta _{ac}\omega _{bd}-\frac{11}{3}\eta _{ab}\omega _{cd}\right) \notag \\
&+&\frac{2}{3}\gamma p^{4}\left( \eta _{ab}\eta _{cd}-2\eta _{ab}\omega _{cd}+\omega _{ab}\omega _{cd}\right)  \notag \\
&-&\frac{1}{4}\kappa v^{2}p^{4}\left(\eta _{ac}\eta _{bd}+\frac{13}{3}\omega _{ab}\omega _{cd}-2\eta
_{ac}\omega _{bd}+\frac{10}{3}\eta _{ab}\eta _{cd}-\frac{20}{3}\eta _{ab}\omega_{cd}\right)  \notag \\
&+&\frac{1}{2}\kappa p^{2}\left( v\cdot p\right) ^{2}\left( \eta _{bc}\eta _{ad}-\eta _{ad}\omega _{bc}\right)+\cdots .  \label{op sem}
\end{eqnarray}
The ellipses in the above expressions refer to terms with the same prior
structure, but switching $a\longleftrightarrow b$, $b\leftrightarrow c$,
$ab\leftrightarrow cd$ in such a way that the wave operator is symmetric
under these exchanges.

The fundamental structures that effectively contribute to the operational
character of $\mathcal{O}^{2}$ correspond to metrics and derivatives or
metrics and momenta in Fourier space. Thus, we expect to expand $\mathcal{O}^{2}$ 
in terms of the Barnes-Rivers operators [Eqs. (\ref{projspin2}) \textemdash (\ref{mapspin021})]. 
In fact, one can show that
\begin{eqnarray}
\mathcal{O}_{ab;cd}^{2} &=&\frac{1}{2}p^{2}\left\{\left( \alpha +\left( \beta -\frac{1}{2}\kappa b^{2}\right)p^{2}\right) P(2)
+\left[ \left(6\beta+2\gamma -\frac{11}{4}\kappa b^{2}\right) p^{2}-\alpha \right]P_{11}(0)\right\} _{ab;cd}  \notag \\
&+&\frac{1}{2}\kappa p^{2}\left( b\cdot p\right) ^{2}\left( P(2)+P_{11}(0)+\frac{1}{2}P(1)\right) _{ab;cd} .  \label{op 2}
\end{eqnarray}

Once the $\mathcal{O}^{1}$ operator contains in its internal structure the
background vector, $b_{a}$, we cannot expand it in terms of the
Barnes-Rivers operators. The direction in spacetime defined by $b_{a}$
breaks Lorentz symmetry, promoting the coupling between the d.o.f.
of the distinct spins defined by Eq. (\ref{projspin2}) \textemdash (\ref{projspin02}). 
In the cases of spins $1$ and $2$, which define subspaces with
dimension $3$ and $5$, respectively, the background vector yields a complete
splitting into one-dimensional subspaces. Furthermore, the structure of the
terms in Eq. (\ref{op quebra}) establishes the possible coupling between these
d.o.f. In order to present the projectors that compose the
Barnes-Rivers operators, we firstly write the background vector in terms of
the momentum and of a spacelike vector as
\begin{equation}
b_{a}=\frac{b\cdot p}{p^{2}}p_{a}+\sqrt{\frac{p_{\ast }^{2}}{p^{2}}}%
e_{a}^{3},
\end{equation}
where 
\begin{equation}
p_{\ast }^{2}=(b\cdot p)^{2}-b^{2}p^{2}  \label{p*}
\end{equation}%
and $e_{a}^{3}$ is a spacelike vector orthogonal to $p$. If we define $e_{1}$
and $e_{2}$ such that
\begin{eqnarray}
e_{i}\cdot e_{j} &=&-\delta _{ij},\ \ \ e_{i}\cdot p=0,  \label{orto1}
\end{eqnarray}
we can split the transverse operator as
\begin{equation}
\theta =\rho +\sigma +\tau,  \label{dec}
\end{equation}
with
\begin{eqnarray}
\rho _{ab} &=&-e_{a}^{1}e_{b}^{1},\ \ \ \sigma _{ab} =-e_{a}^{2}e_{b}^{2},\ \ \ \tau _{ab} =-e_{a}^{3}e_{b}^{3}.  \label{ro sigma tau}
\end{eqnarray}

With these definitions, we can split the Barnes-Rivers operators in terms of
projectors which act in each one of the spin subspaces. Also, we can define
mappers among the subspaces defined by these projectors. Any operator,
projector, or mapper can be written, up to changes of basis by unitary
transformations, as
\begin{equation}
P\left( IJ\right) _{\left( ij\right) }=\left( -\right) ^{R+S}\psi \left(
I\right) _{\left( i\right) }\psi \left( J\right) _{\left( j\right) },
\label{carlos1}
\end{equation}
where $\psi \left( I\right) _{\left( i\right) }$ is the $i$th eigenvector of
the Barnes-Rivers spin projector $P\left( I\right)$. $R$, $S$ can assume
the values $0$ and $1$, corresponding to the spin parities $+1$ or $-1$,
respectively. By convention, we choose the subscripts $i$ and $j$ to range
from $1$ to $5$ for the spin $2$ $\left( I,J=2\right) $, to assume the values $6$
and $7$ for the $0$-spins $\left( I,J=0\right) $ defined by Eqs. (\ref{projspin01})
and (\ref{projspin02}), respectively, and to range from $8$ to $10$ for the spin
$1$ $\left( I,J=1\right) $. In terms of the vectors in Eq. (\ref{orto1}), we can write
the $\psi $ eigenvectors as
\begin{eqnarray}
\psi(2)_{(1)ab}&=&\frac{1}{\sqrt{2}}\left(e_{1a}e_{2b}+e_{2a}e_{1b}\right)
\label{psi 21}, \\
\psi(2)_{(2)ab}&=&\frac{1}{\sqrt{2}}\left(e_{1a}e_{3b}+e_{3a}e_{1b}\right), \\
\psi(2)_{(3)ab}&=&\frac{1}{\sqrt{2}}\left(e_{2a}e_{3b}+e_{3a}e_{2b}\right), \\
\psi(2)_{(4)ab}&=&\frac{1}{\sqrt{2}}\left( \rho _{ab}-\sigma_{ab}\right), \\
\psi(2)_{(5)ab}&=&\frac{1}{\sqrt{6}}\left( \rho _{ab}+\sigma
_{ab}-2\tau_{ab}\right), \\
\psi(0)_{(6)ab}&=&\frac{1}{\sqrt{3}}\theta _{ab}, \\
\psi(0)_{(7)ab}&=&\omega _{ab}, \\
\psi(1)_{(8)ab}&=&-\frac{1}{\sqrt{2}}\left( e_{a}^{1}\frac{p_{b}}{\sqrt{p^{2}%
}}+e_{b}^{1}\frac{p_{a}}{\sqrt{p^{2}}}\right), \\
\psi(1)_{(9)ab}&=&-\frac{1}{\sqrt{2}}\left(e_{a}^{2}\frac{p_{b}}{\sqrt{p^{2}}%
}+e_{b}^{2}\frac{p_{a}}{\sqrt{p^{2}}}\right), \\
\psi(1)_{(10)ab}&=&-\frac{1}{\sqrt{2}}\left(e_{a}^{3}\frac{p_{b}}{\sqrt{p^{2}%
}}+e_{b}^{3}\frac{p_{a}}{\sqrt{p^{2}}}\right).  \label{psi 110}
\end{eqnarray}

Using Eq. (\ref{orto1}), the normalization of the $\psi$ eigenvectors is
given by
\begin{equation}
\psi \left( I\right) _{\left( i\right) ab}\psi \left( J\right) _{\left(
j\right) }^{ab}=\left( -\right) ^{P}\delta ^{IJ}\delta _{ij}.
\end{equation}

The operators of the type $P_{ij}\left( JM\right)$ are objects that map the
subspace defined by $P_{ii}\left( JJ\right)$ in the subspace defined by 
$P_{jj}\left( MM\right) $. The algebra that these operators satisfy is
orthonormal and complete in the following sense:
\begin{eqnarray}
P_{ij}\left( JM\right) _{ab;fg}P_{kl}\left(NP\right) _{\ \;cd}^{fg} &=&\delta _{jk}\delta
_{MN}P_{il}\left(JP\right) _{ab;cd},  \label{orto} \\
{{\sum_{i,J}}}P_{ii}(JJ) &=&1.
\label{comp}
\end{eqnarray}

In terms of these operators, we can write $\mathcal{O}^{1}$ as follows:
\begin{eqnarray}
\mathcal{O}^{1} &=&\frac{1}{2}\kappa p^{2}\Big[\frac{1}{2}\left( \left( b\cdot p\right)
^{2}+b^{2}p^{2}\right) \left( P_{22}(2-2)+P_{33}(2-2)\right)  \notag \\
&+&\frac{1}{3}\left( \left( b\cdot p\right)^{2}+2b^{2}p^{2}\right) P_{55}(2-2)+\left( b\cdot p\right) ^{2}( P_{11}(2-2)\notag \\
&+&P_{44}(2-2))+\frac{1}{3}\left( 2\left(b\cdot p\right) ^{2}+b^{2}p^{2}\right) P_{11}(0-0)  \notag \\
&+&\frac{\sqrt{2}}{3}( \left( b\cdot p\right)^{2}-b^{2}p^{2}) \left( P_{15}(0-2)+P_{51}(2-0)\right)\Big].
\label{op quebra spin}
\end{eqnarray}

The expansion of the wave operator in terms of the operators in Eq. (\ref{orto})
presents, in general, coefficients $a\left( JM\right)$ organized in matrix
blocks that show a coupling between the $J$ and $M$ spins. To avoid
redundancy in the notation, we denote the coefficient matrices associated
with operators of the $J$ and $M$ spins simply as $a\left( JM\right)$, with a
random order in the appearance of the $J$ and $M$ letters. Thus, the $a(JJ)$, 
$a(MM)$, and $a(MJ)$ coefficients are parts of the same matrix $a(JM)$ 
in such a way that a wave operator is generally written as
\begin{equation}
\mathcal{O}=\sum_{ij}a_{ij}\left( JM\right) P_{ij}\left( JM\right).
\label{gen op}
\end{equation}
The sum over the $JM$ indices is already contained in the sum over the $ij$ ones,
according to our notation in Eqs. (\ref{psi 21}) \textemdash (\ref{psi 110}).

In the case where the matrices $a(JM)$ are invertible, the propagator
saturated with the emission and absorption sources of particles is given by
\begin{equation}
\Pi =i\sum_{ij}a^{-1}\left( JM\right) _{ij}J^{\ast ab}P_{ij}\left( JM\right)
_{ab;cd}J^{cd}.  \label{prop saturado}
\end{equation}
However, as already discussed, the Lagrangian of Eq. (\ref{perturba}) has the gauge
symmetry of Eq. (\ref{gauge}), which implies the absence of spin-$1$ and spin-$0$
sectors, as defined by Eqs. (\ref{projspin1}) and (\ref{projspin02}),
respectively. In this case, the wave operator is not invertible. Nevertheless,
the gauge invariance also requires that the sources of emission and
absorption satisfy conditions such that there is no emission and
absorption of the modes that depend on the gauge. Thus, the contractions of
the spin-$1$ and spin-$0$ operators [Eqs. (\ref{projspin1}) and (\ref{projspin02})]
with the sources should be null, resulting in the gauge invariant propagator
given by Eq. (\ref{prop saturado}) with only the inverses of the largest
nondegenerate submatrices.

In our model, the complete wave operator [Eq. (\ref{gen op})] presents only a
nonvanishing matrix, that is given by
\begin{equation}
a\left( 02\right) =\left( 
\begin{array}{cccccc}
A & 0 & 0 & 0 & 0 & 0 \\ 
0 & B & 0 & 0 & 0 & 0 \\ 
0 & 0 & B & 0 & 0 & 0 \\ 
0 & 0 & 0 & A & 0 & 0 \\ 
0 & 0 & 0 & 0 & C & F \\ 
0 & 0 & 0 & 0 & F & D%
\end{array}%
\right) ,  \label{coef1}
\end{equation}
with
\begin{eqnarray}
A &=&\frac{1}{2}p^{2}\left( \alpha
+\beta p^{2}+\kappa \left( \left( b\cdot p\right) ^{2}-\frac{1}{2}
b^{2}p^{2}\right) \right) ,  \label{coefa} \\
B &=&\frac{1}{2}p^{2}\left( \alpha
+\beta p^{2}+\frac{1}{2}\kappa \left( b\cdot p\right) ^{2}\right) ,
\label{coefb} \\
C &=&\frac{1}{2}p^{2}\left( \alpha
+\beta p^{2}+\frac{1}{3}\kappa \left( \left( b\cdot p\right) ^{2}+\frac{1}{2}%
b^{2}p^{2}\right) \right) ,  \label{coefc} \\
D &=&p^{2}\left[ \left( 6\beta +2\gamma
\right) p^{2}-\alpha \right] +\frac{1}{6}\kappa p^{2}\left( 2\left( b\cdot
p\right) ^{2}-\frac{31}{2}b^{2}p^{2}\right) ,  \label{coefd} \\
F &=&\frac{\sqrt{2}}{6}\kappa
p^{2}\left( \left( b\cdot p\right) ^{2}-b^{2}p^{2}\right) .  \label{coeff}
\end{eqnarray}

The constraints satisfied by the sources are given by
\begin{eqnarray}
p_{a}\mathcal{J}^{ab} &=&p_{a}\mathcal{J}^{ba}=0,  \\
\tau _{ab}\mathcal{J}^{bc} &=&\tau _{ab}\mathcal{J}^{cb}=0,
\label{vinculo fontes}
\end{eqnarray}
where the last condition is valid only on the mass shell. These identities
are responsible for the inhibition of the spin-$1$ and spin-$0$ modes
associated with the gauge invariance of the model.

As we can see, the matrix $a(02)$ assumes a block-diagonal form. The lower-right 
block points to a coupling between the spin $0$ and the fifth
component of the spin $2$. The other components of the spin $2$ do not couple
to the other spins. However, the distinction of the coefficients, $A$
and $B$, indicates that the propagation of components $1$ and $4$ have 
dynamics independent of the components $2$ and $3$. This splitting of the
spin-$2$ matrix into direct-sum-of-$U\left( 1\right)$ complex-conjugate-related 
representations is compatible with the CPT invariance and the
breaking of the Lorentz symmetry by a background vector. The structure of
the $a(02)$ matrix suggests that we can split it into three distinct blocks:
\begin{equation}
a^{14}\left( 2\right) =\left( 
\begin{array}{cc}
A & 0 \\ 
0 & A%
\end{array}%
\right) ,  \label{matriz14}
\end{equation}
\begin{equation}
a^{23}\left( 2\right) =\left( 
\begin{array}{cc}
B & 0 \\ 
0 & B%
\end{array}%
\right) ,  \label{matriz23}
\end{equation}
\begin{equation}
a^{56}\left( 02\right) =\left( 
\begin{array}{cc}
C & F \\ 
F & D%
\end{array}%
\right) .  \label{matriz56}
\end{equation}

The matrices $a^{14}$ and $a^{23}$ only carry information about the dynamics
of the spin $2$, while the matrix $a^{56}$ shows the conjunction of the spin-$0$
d.o.f. with one of the d.o.f. of the spin $2$.

\subsection{Analysis of the particle content \label{subsection 2.2}}

To analyze whether the particle content described by the model in Eq. (\ref{onda
quebra}) is tachyon and ghost free, we explore the structure of the
propagator [Eq. (\ref{prop saturado})] expressed in terms of the inverses of the
matrices in Eqs. (\ref{matriz14})\textemdash(\ref{matriz56}). The saturated propagator has
poles when the determinants of these matrices vanish. In a Lorentz-invariant model, the zeros of these determinants
always occur for values of $p^{2}$ such that $p^{2}=m^{2}$, with $m^{2}$
being defined as the particle mass. The condition for the absence of
tachyons in this case is given by $m^{2}>0$. This condition assures positive-energy modes for any
value of the momentum $\vec{p}$, preventing the arising of instabilities due
to the absence of a minimum energy.

For LV, the poles of the propagators may
take forms more general than $p^{2}=m^{2}$, due to the presence of the
background tensors. In this case, the discussion of stability is a much more
subtle issue. The positivity of energy for all $\vec{p}$ and the observer Lorentz
invariance are insufficient to prevent the appearance of spacelike
momenta in the dispersion relations. In a strongly boosted frame, this
implies the reemergence of negative energies, spoiling the supposed stability
of the model. So, besides requiring the positivity of the energy for all $\vec{p}$
in one frame, one must also impose that spacelike momenta are absent
or, equivalently, that the energy is positive in all frames. 

For a general quadratic form built from the energy and momentum, these
requirements imply that the particle and antiparticle energy solutions of
the dispersion relations should correspond to positive and negative branches
of a hyperbola whose asymptotes lie inside or on the light cone in the Minkowski
causal diagram.

In spite of the possible solution to the stability problem
in the way we have described, this pattern of dispersion relation raises
other problems related to microcausality \cite{alan-lehnert,klink,reyes,potting}. In fact, as discussed in Ref.
\cite{alan-lehnert}, the microcausality condition, that expresses that fields
should commute for spacelike separations of their arguments, imposes that
the modulus of the group velocity of the wave packets, formed out from the
superposition of plane waves that satisfy the associated dispersion
relations, must be lower than or at most equal to $1$ (in natural units). But this
requirement, in general, clashes with the stability constraint,
since there, the slopes of the asymptotes are the limit of the group
velocity and, as discussed above, should be greater than or at least equal to $1$.
Apparently, the only exception is the limiting situation where the asymptotes
lie exactly on the light cone.

The mentioned difficulty in simultaneously attaining stability and causality
in LV models originates in the role that these assumptions
play in the search for a Lorentz-invariant description of quantum theory.
For the moment, we must only discuss the conditions of stability and
unitarity and shall comment where the microcausal problem can appear. A
detailed investigation of this question should be considered, taking into account the
role of the extra modes arising from the SSB mechanism responsible for the
restoration of the Lorentz symmetry.

Another criterion analyzed in this work concerns the probabilistic character
of processes in QFT. The unitarity of the scattering S matrix
is a compulsory requirement to achieve any reasonable interpretation of the
results of scattering processes. Since the residue of the propagator
evaluated at the poles provides information on the norms of the states
associated with the propagating mode, we impose that these norms must be
positive definite. Considering what was said about stability and unitarity,
our investigation in this work concerns the implementation of the following
conditions for the absence of tachyons and ghosts:
\begin{equation}
p_{0}=f\left( \vec{p},m^{2}\right) >0,\ \ \ \ p^{2}\geq 0\ \ \forall \ \ 
\vec{p},  \label{tachyon}
\end{equation}
\begin{equation}
\mathcal{I}Res\left( \Pi \big|_{p_{0}=f\left( \vec{p}\right) }\right) >0,
\label{ghost}
\end{equation}%
where $f\left( \vec{p},m^{2}\right)$ is an arbitrary function of the momenta
with the masses of modes, $m^{2}$, depending on the constants of the model,
including the background vector. These functions are obtained as roots of
the polynomials defined by the determinants of matrices the in Eqs. (\ref{matriz14})\textemdash(\ref{matriz56}).

Using Eq. (\ref{carlos1}), we can rewrite the propagator as
\begin{equation}
\Pi =i\left( -\right) ^{R+S}{{\sum_{ij}}}S_{i}^{\ast
}A_{ij}^{-1}\left( JM,m^{2}\right) S_{j}\left( p_{0}-f\left( \vec{p}
,m^{2}\right) \right) ^{-1},  \label{prop modo}
\end{equation}
where $S_{i}=\psi _{i}^{ab}J_{ab}$ and $A_{ij}^{-1}\left( JM,m^{2}\right)$
is the matrix $a\left( JM\right) ^{-1}$ with the pole $f\left( \vec{p},m^{2}\right)$ 
extracted. Thus, the positivity condition [Eq. (\ref{ghost})] for
arbitrary sources is guaranteed if the eigenvalues of $A_{ij}^{-1}\left(JM,m^{2}\right)$ 
are positive (negative) defined for the case when $R+S=0$ $(1)$. Moreover, 
one can show that $A_{ij}^{-1}\left( JM,m^{2}\right) $ has
only one non-null eigenvalue whenever evaluated at the pole, so that a
positive (negative) eigenvalue is ensured by the positivity (negativity) of
the trace $A_{ij}^{-1}\left( JM,m^{2}\right)$ evaluated at the pole. We can
then rewrite the condition for the absence of ghosts as
\begin{equation}
\left( -1\right) ^{R+S}trA_{ij}^{-1}\left( JM,m^{2}\right) |_{p_{0}=f\left( 
\vec{p},m^{2}\right) }>0.  \label{traco}
\end{equation}
Taking the inverse of the matrices in Eqs. (\ref{matriz14})\textemdash(\ref{matriz56}), we can
see that each one of them presents a simple massless pole ($p^{2}=0$) besides
simple massive poles in the $14$ and $23$ sectors and a double massive pole
in the $56$ sector. The massive poles, which appear in matrices of the $14$ and 
$23$ sectors, can be put in the following general quadratic form:
\begin{equation}
A_{ij}p_{0}^{2}+B_{ij}\vec{p}^{2}+C_{ij}|\vec{p}|p_{0}+D_{ij}=0, \label{dispersion 14-23}
\end{equation}
where each pair of subindeces $(i,j)$ can assume the values $(1,4)$ and $(2,3)$,
referring to the respective alluded sectors. These sets of parameters are
given by 
\begin{eqnarray}
A_{14} &=&\beta +\frac{1}{2}\kappa \left( b_{0}^{2}+\vec{b}^{2}\right), \label{14 param1} \\
B_{14} &=&-\left( \beta -\frac{1}{2}\kappa \left( b_{0}^{2}-\vec{b}^{2}\right) -\kappa \vec{b}^{2}\cos ^{2}\theta \right), \\
C_{14} &=&-2\kappa b_{0}\left\vert \vec{b}\right\vert \cos \theta , \\
D_{14} &=&\alpha  \label{14 param2},\\
A_{23} &=&\beta +\frac{1}{2}\kappa b_{0}^{2}, \label{23 param1}\\
B_{23} &=&-\beta +\frac{1}{2}\kappa \vec{b}^{2}\cos ^{2}\theta , \\
C_{23} &=&-\kappa b_{0}\left\vert \vec{p}\right\vert \cos \theta , \\
D_{23} &=&\alpha.   \label{23 param2}
\end{eqnarray}

For general values of the parameters, Eq. (\ref{dispersion 14-23})
may describe ellipses, parabolas, or hyperbolas on the plane ($p_{0}$,$|\vec{p}|$) 
with the center at the origin or any of its degenerate situations.
For physical reasons, and in agreement with the condition for the absence of
tachyons, we demand that it have two real roots for $p_{0}$ and that each one of
these roots have single-valued positivity. Following our previous
considerations, the positivity of energy is ensured in all reference frames
if spacelike four-momenta are avoided. This can be reached by imposing the
existence of asymptotes constrained to lie inside the light cone. This is
only possible for the hyperbolic case. In terms of the parameters of the
quadratic form, we have
\begin{eqnarray}
C_{ij}^{2}-4A_{ij}B_{ij} &>&0,  \label{cond asymp} \\
\left\vert m_{\pm }\right\vert  &=&\left\vert -\frac{C_{ij}}{2A_{ij}}\pm 
\sqrt{\left( \frac{C_{ij}}{2A_{ij}}\right) -\frac{D_{ij}}{A_{ij}}}
\right\vert \geq 1, \\
D_{ij}/A_{ij} &<&0,  \label{cond asymp 2}
\end{eqnarray}
where $m_{\pm }$ are the slopes of the two asymptotes of the hyperbola.

The poles of the $56$ sector are the solutions of the quartic equation of
the form 
\begin{equation}
\lambda _{1}p_{0}^{4}+\lambda _{2}\vec{p}^{4}+\lambda _{3}|\vec{p}
|^{3}p_{0}+\lambda _{4}|\vec{p}|p_{0}^{3}+\lambda _{5}p_{0}^{2}+\lambda _{6}
\vec{p}^{2}+\lambda _{7}\vec{p}^{2}p_{0}^{2}+\lambda _{8}=0, \label{dispersion 56}
\end{equation}
where the coefficients ($\lambda _{1},\ldots ,\lambda _{8}$) depend on the
Lagrangian parameters and on the angle $\theta$ between the vector $\vec{b}$
and the particle momentum $\vec{p}$. In this case, there are many other
possible situations for the curves in the plane ($p_{0}$,$|\vec{p}|$) as
compared with the previous one. Obviously, there are physically reasonable
possibilities among them. As a simple example, we have the interesting
situation in which the quartic form is separable in the product of two
quadratic ones. Within this decomposition, we can demand that each one
correspond to a hyperbola with the same desired properties of the previous
discussion. Due to the complexity of analyzing this dispersion relation in
general, let us concentrate on the other sectors and return to this case
when we discuss the unitarity of the model.

By inspecting the conditions in Eqs. (\ref{cond asymp})\textemdash(\ref{cond asymp 2}) and using the
two sets of parameters in Eqs. (\ref{14 param1})\textemdash(\ref{14 param2}) and Eqs. (\ref{23 param1})\textemdash(\ref{23 param2}), 
we conclude that the parameters of the Lagrangian [Eq. (\ref{principal})] must satisfy the conditions 
\begin{eqnarray}
14\ \ \text{Sector} &:&\kappa >0,\ \ \beta <0\ \ b^{2}>-2\left\vert \frac{\beta }{\kappa }\right\vert ,\ \ \left( b_{0}^{2}+\vec{b}^{2}\right) <2\left\vert 
\frac{\beta }{\kappa }\right\vert ,\ \ \alpha >0,  \label{1cond tac 14} \\
&&\kappa <0,\ \ \beta >0\ \ b^{2}>-2\left\vert \frac{\beta }{\kappa }%
\right\vert ,\ \ \left( b_{0}^{2}+\vec{b}^{2}\right) <2\left\vert \frac{%
\beta }{\kappa }\right\vert ,\ \ \alpha <0, \label{2cond tac 14}\\
23\ \ \text{Sector} &:&\kappa >0,\ \ \beta <0,\ \ b_{0}^{2}<2\left\vert \frac{\beta }{%
\kappa }\right\vert ,\ \ \alpha >0,  \label{1cond tac 23} \\
&&\kappa <0,\ \ \beta >0,\ \ b_{0}^{2}<2\left\vert \frac{\beta }{\kappa }%
\right\vert ,\ \ \alpha <0,  \label{2cond tac 23}
\end{eqnarray}
in order to avoid instabilities coming from these sectors. It is worthwhile
emphasizing that the same set of conditions is required for any choice of
the background vector: $b^{2}>0$, $b^{2}<0$, or $b^{2}=0$.
 
We can assume the Sun-centered reference frame (SCRF)
to be the one where the components of the background vector are sufficiently
small in such a way as to match the many experiments that are in agreement with
Lorentz invariance. In this sense, the relations in Eqs. (\ref{1cond tac 14})\textemdash(\ref{2cond tac 23}) 
impose limits on the possible boosted reference frames
related to the SCRF. Apparently, this dependence of the physical consistency on some restricted set
of inertial observers clashes with the observer invariance of the starting
Lagrangian. Nevertheless, one should be aware that our discussion about physical consistency
relies on the validity of the perturbative approach to QFT and, for enough strongly boosted frames,
the LV parameters are not supposed to be small anymore, jeopardizing the conclusions obtained with perturbation theory.

Let us turn to the discussion of the fulfillment of the condition in Eq. (\ref{ghost}). 
As we already stated, this is needed in order to get a unitary
model. For the massless pole, we can use the source constraints [Eq. (\ref{vinculo fontes})], 
the inverse of matrices (\ref{matriz14})-(\ref{matriz56}),
and the condition for the absence of ghosts [Eq. (\ref{ghost})] at the pole $p^{2}=0$ to get
\begin{equation}
J^{\ast ab}\frac{1}{2\left( \alpha +\kappa \left( v\cdot
p\right) ^{2}\right) }\Bigg[\left( \eta _{ac}\eta _{bd}+\eta _{ad}\eta
_{bc}\right) -\eta _{ab}\eta _{cd}\Bigg]\Bigg|_{p^{2}=0}J^{cd}>0.  \label{prop grav}
\end{equation}

We recognize the structure inside the brackets as the residue matrix of the
graviton propagator. In comparison with the EH model, we have the
appearance of the $\kappa \left( v\cdot p\right) ^{2}$ term following the 
Newton constant $\alpha$. This contribution to the Newton constant, due to
the LV term, denotes a better UV behavior, at least on the
mass-shell analysis, and could have interesting consequences in the searches
for renormalizable gravity models.

To assure the positivity of the residue of the propagator for the massless
mode [Eq. (\ref{prop grav})], we must impose
\begin{equation}
\left( \alpha +\kappa \left( v\cdot p\right) ^{2}\right) >0.
\end{equation}
For any type of background vector, this condition is satisfied for all $\vec{p}$ if we require that
\begin{eqnarray}
\kappa  &>&0,  \label{cond kapa n mass} \\
\alpha  &>&0.  \label{cond alfa n mas}
\end{eqnarray}
By reanalysing the conditions for stability given in Eqs. (\ref{1cond tac 14})\textemdash(\ref{2cond tac 23}), 
we conclude that only the conditions in Eqs. (\ref{1cond tac 14}) and (\ref{2cond tac 23}) are compatible with these new constraints. One
important point to notice is that, even if we relax the constraint
on the positivity of $\kappa $ by requiring the positivity of the
combination $\left[ \alpha +\kappa \left( v\cdot p\right) ^{2}\right]$ up
to some momentum cutoff, we still have problems with the conditions in Eqs. (\ref{2cond tac 14}) 
and (\ref{2cond tac 23}), since they require $\alpha <0$,
whereas low-energy unitarity requires $\alpha >0$.

For the massive poles of the $ij$ sectors, the condition on the residues of
the propagators demands that
\begin{equation}
\left( p_{0}-f_{ij}^{n}\left( \vec{p}\right) \right) tra_{ij}^{-1}\Big|
_{p_{0}=f_{ij}^{n}}>0, \label{cond ghost ij}
\end{equation}
where $f_{ij}^{n}$ means the $n$th positive energy root of the polynomial
dispersion relation of the $ij$ sector. Only for the $56$ sector can the $n$ index 
be nontrivial, since there can be two positive energy solutions for the
fourth-degree polynomial [Eq. (\ref{dispersion 56})]. Solving Eq. (\ref{dispersion 14-23}) 
with the two sets of parameters [Eqs. (\ref{14 param1})\textemdash(\ref{23 param2})] and imposing the conditions 
in Eqs. (\ref{cond asymp})\textemdash(\ref{cond asymp 2}), we get the two positive energy solutions
{\small {\ 
\begin{eqnarray}
f_{14}&=&\frac{\kappa b_{0}\left(\vec{b}\cdot\vec{p}\right)}{\beta+\frac{1}{2%
}\kappa\left(b_{0}^{2}+\vec{b}^{2}\right)}+\left(\left(\frac{\kappa
b_{0}\left(\vec{b}\cdot\vec{p}\right)}{\beta+\frac{1}{2}\kappa%
\left(b_{0}^{2}+\vec{b}^{2}\right)}\right)^{2}+\frac{\left(\beta-\frac{1}{2}%
\kappa b^{2}\right)\vec{p}^{2}-\kappa\left(\vec{b}\cdot\vec{p}%
\right)^{2}-\alpha}{\beta+\frac{1}{2}\kappa\left(b_{0}^{2}+\vec{b}^{2}\right)%
}\right)^{1/2}, \\
f_{23}&=&\frac{\kappa b_{0}\left(\vec{b}\cdot\vec{p}\right)}{2\left(\beta+%
\frac{1}{2}\kappa b_{0}^{2}\right)}+\left(\left(\frac{\kappa b_{0}\left(\vec{%
b}\cdot\vec{p}\right)}{2\left(\beta+\frac{1}{2}\kappa b_{0}^{2}\right)}%
\right)^{2}+\frac{\beta\vec{p}^{2}-\frac{1}{2}\kappa\left(\vec{b}\cdot\vec{p}%
\right)^{2}-\alpha}{\beta+\frac{1}{2}\kappa b_{0}^{2}}\right)^{1/2}.
\end{eqnarray}
} }

Taking the inverse of the matrices (\ref{matriz14}) and (\ref{matriz23})
with the coefficients in Eqs. (\ref{coefa})-(\ref{coeff}) and using the condition 
in Eq. (\ref{cond ghost ij}), we get the following inequalities for the $14$ and 
$23$ sectors, respectively:

{\small {\ 
\begin{eqnarray}
2\left( \kappa b_{0}\left( \vec{b}\cdot \vec{p}\right) \right) \left( \left( 
\frac{\kappa b_{0}\left( \vec{b}\cdot \vec{p}\right) }{2\left( \beta +\frac{1%
}{2}\kappa b_{0}^{2}\right) }\right) ^{2}+\frac{\beta \vec{p}^{2}-\frac{1}{2}%
\kappa \left( \vec{b}\cdot \vec{p}\right) ^{2}-\alpha }{\beta +\frac{1}{2}%
\kappa b_{0}^{2}}\right) ^{1/2} > \\
-2\frac{\left( \kappa b_{0}\left( \vec{b}\cdot \vec{p}\right) \right) ^{2}%
}{\beta +\frac{1}{2}\kappa \left( b_{0}^{2}+\vec{b}{}^{2}\right) }+\kappa
b_{0}^{2}\vec{p}^{2}+\kappa \left( \vec{b}\cdot \vec{p}\right) ^{2}+\alpha ,
\notag  \label{ghost cond 14} \\
\left( \kappa b_{0}\left( \vec{b}\cdot \vec{p}\right) \right) \left( \left( 
\frac{\kappa b_{0}\left( \vec{b}\cdot \vec{p}\right) }{2\left( \beta +\frac{1%
}{2}\kappa b_{0}^{2}\right) }\right) ^{2}+\frac{\beta \vec{p}^{2}-\frac{1}{2}%
\kappa \left( \vec{b}\cdot \vec{p}\right) ^{2}-\alpha }{\beta +\frac{1}{2}%
\kappa b_{0}^{2}}\right) ^{1/2} > \\
-\frac{\left( \kappa b_{0}\left( \vec{b}\cdot \vec{p}\right) \right) ^{2}}{%
2\left( \beta +\frac{1}{2}\kappa b_{0}^{2}\right) }+\frac{1}{2}\kappa \left( 
\vec{b}\cdot \vec{p}\right) ^{2}+\alpha ^{2}+\frac{1}{2}\kappa b_{0}^{2}\vec{%
p}^{2}.  \notag  \label{ghost cond 23}
\end{eqnarray}%
} }From the condition in Eq. (\ref{1cond tac 14}), we see that $\alpha >0$, $\beta<0$, 
$\kappa >0$, and $\beta +\frac{1}{2}\kappa \left( b_{0}^{2}+\vec{b}^{2}\right) <0$. 
So, the right-hand side of expression (\ref{ghost cond 14})
is positive definite, and so should be the left-hand side, but this is false,
since $b_{0}$ and $\vec{b}\cdot \vec{p}$ have no definite sign. The same
reasoning may be applied to the expression (\ref{ghost cond 23}) with the
aid of the conditions in Eq. (\ref{1cond tac 23}), and we again reach a contradiction.

The conclusion of this analysis is that the propagation of these modes
violates the unitarity of the S matrix. The only way to try to fix this problem
is inhibiting the propagation of these modes. However, if we look for the
quadratic form [Eq. (\ref{dispersion 14-23})], we note that to force the
absence of these modes in any reference frame we must impose that 
$A=B=C=0$, and from Eqs. (\ref{14 param1})\textemdash(\ref{23 param2}) we see that this is
only possible if $\kappa =0$ and $\beta =0$. That is, the LV
term in the Lagrangian [Eq. (\ref{principal})] brings unavoidable inconsistencies
to the quantum perturbative description of the model expanded around the
flat Minkowski metric and must be switched off in order to give rise a
healthy particle spectrum.

With this considerable change in the scenario, the $56$ sector becomes
approachable. By making $\kappa =\beta =0,$ the $0$ component of the spin $2$
decouples from the spin $0$ in the matrix [Eq. (\ref{coef1})], and the other
coefficients become equal ($A=B=C=D$), setting up the coefficient of the
spin-$2$ projector. Furthermore, the expression (\ref{dispersion 56}) becomes
\begin{eqnarray}
\alpha \gamma p^{2}-\frac{1}{2}\alpha ^{2} &=&0.  \label{disp 56 k=0}
\end{eqnarray}
Considering our previous conditions for unitarity and stability, we must
impose that $\gamma >0$ in order to get a stable spin-$0$ mode. The residue
matrix becomes only $\frac{1}{2\gamma p^{2}}$, and the positivity of $\gamma$
also ensures that this is a nonghost mode. With these choices, the model
with Lagrangian $\alpha R+\gamma R^{2}$ propagates two modes: a massless
spin $2$, identified as the graviton, and a massive spin $0$. This is a well-known model in the context of higher-derivative gravity \cite{nieu}. In spite of the
presence of the higher derivatives, the model must be complemented with another $\left(R_{\mu\nu}R^{\mu\nu}\right)^{2}$ 
term to attain renormalizability, but as we have discussed in the
Introduction, this insertion spoils the unitarity.

\section{Discussion and Conclusions \label{section 3}}

We started with a general higher-derivative gravity model added with a
LV CPT-even term also containing higher derivatives, since we
intended to analyze the role of the Lorentz-invariance assumption in the
incompatibility of renormalizability and unitarity of higher-derivative gravity models. Our
guide to propose such a term was grounded on the general framework of the
Standard-Model Extension, where this kind of contribution arises as a
spontaneous breaking of the Lorentz symmetry at some high energy in a
fundamental theory. The assumption of spontaneous instead of explicit
symmetry breaking is particularly mandatory in the gravity sector of the SME if
we do not want to abandon the most usual geometrical interpretation of the
gravitational interaction. The most striking signal of spontaneous symmetry
breaking is the appearance of extra massless and massive modes as a result of
the breaking. Our position was only to discuss the metric modes by hoping
that this situation would correspond to scenarios where the interaction
that promotes the coupling between the metric and extra modes does not
affect the quadratic Lagrangian. 

To our mind, this assumption should be further investigated, since the
extra modes are responsible for making the whole model consistent. An
explicit example of a potential triggering the spontaneous breaking of the
Lorentz symmetry and its complete analysis would shed some light on this
question and also could change our conclusions. Another point to be
highlighted is that, as an effective model, the SME only aims to provide low-energy
effects coming from some more fundamental theory, and as such, it cannot be
pushed to extremely high energies without expecting that some inconsistencies will
appear. In our investigation, we tried to discuss Lorentz violation
independent of this restriction, but without clashing with it, and our
general considerations of stability and unitarity point to the conclusion
that the extra Lorentz noninvariant term is unable to fix the mentioned
problem of the incompatibility of renormalizability and unitarity. Within the
philosophy of effective field theories, the terms considered here can be
accepted if we constrain the masses of the ghost particles to be of the same
order of the cutoff energy of the model, and the new d.o.f of a
more fundamental theory should be able to fix the situation. However, an
effective discussion of the model should contemplate other possibilities
for the violating terms that respect the same structure as the one considered
here, and it would also be interesting to investigate them.

Acknowledgments

We would like to express our thanks to V.A. Kosteleck\'{y} for useful
discussions. C.H. wishes also to thank the International Center for Spacetime
Symmetries for their kind hospitality. This work has been supported by CAPES
(Coordena\c{c}\~{a}o de Aperfei\c{c}oamento de Pessoal de N\'{\i}vel
Superior-Brazil) and CNPq (Conselho Nacional de Desenvolvimento Cient\'{\i}fico e Tecnol\'{o}gico, Brazil).

\end{document}